\documentclass{kluwer}    
\input psfig.tex
\newdisplay{guess}{Conjecture}

\begin{document}                                                                                   
\begin{article}
\begin{opening}         
\title{Advective Flow Paradigm And Microquasar GRS1915+105} 
\author{Sandip K. \surname{Chakrabarti}}  
\runningauthor{Sandip K. Chakrabarti}
\runningtitle{Advective Flow Paradigm and GRS 1915+105}
\institute{SNBNCBS, JD Block, Salt Lake, Calcutta 700098}
\date{October 15, 2000}

\begin{abstract}
Advective accretion disks and winds are the most self-consistent
solutions today. We describe this paradigm briefly and show how
it attempts to explain some of the interesting observations of the
galactic microquasar GRS1915+105.
\end{abstract}

\end{opening}           

\def\doubarow{\lower-0.05ex\hbox{\rightarrow}\kern-0.55em{\lower0.15ex\hbox{\leftarrow}}}

\noindent To be published in the 3rd Microquasar Workshop, Granada, (Ed.) A.J. Castro-Tirado,
J. Greiner, J. Paredes

\section{Introduction}  

The standard accretion disk model of Shakura and Sunyaev (1973) was
modified immediately after it was proposed. The spectrum
of black hole candidate Cyg X-1 showed (Sunyaev and Tr\"umper, 
1979) that apart from the modified cool black body spectrum, it also 
exhibits a power-law emission at much higher energies, indicating 
the presence of electrons much hotter than a standard disk. Subsequently, 
Sunyaev and Titarchuk (1980) showed that the power-law emission 
is the result of the Comptonization of soft photons from a standard disk
by hot electrons whose source was thought to be {\it outside} the standard disk,
be it in the form of floating `Compton clouds'  or `magnetic corona'.

Meanwhile, transonic flow models were developed in the early eighties 
(Muchotrzeb and Paczy\'nski, 1982, Matsumoto et al. 1984) to modify 
inner edge of the standard disk to ensure that it passes through 
a sonic point.  These solutions ensure that the specific energy of the flow 
remains negative as in a Keplerian disk. However, there is another 
class of transonic disk solutions which are the generalizations of the 
classical Bondi (1952) solutions when angular momentum, viscosity,  
heating, cooling etc. were taken into account. These are known as 
advective flow solutions (Chakrabarti, 1990, 1996a, 1998ab, 2000a). 
Particularly interesting is that these solutions did away with the
external Compton clouds as the inner sub-Keplerian region itself 
is found to behave like one for all practical purposes. 

In this review, I briefly present these solutions for viscous and non-viscous 
flows. The solutions can be suitably combined to obtain a paradigm
of the black hole astrophysics. Truly speaking, a cool, Keplerian disk with a
negative energy cannot smoothly join to most of these {\it steady} solutions
which require positive energy (at least in quasi-adiabatic
flows). For time-dependent
flows such restrictions do not apply. In fact, steady transition from a Keplerian
disk to an advective disk may require systematic heating of a
Keplerian disk or the assumption that all disks are fundamentally
non-steady, only the time scale varies. We also try to explain some
of the observed features of the galactic micro-quasar GRS1915+105.
We find that it always helps to keep a paradigm in the back of one's
mind while attempting to explain any observations. 
Otherwise, each observational feature would demand a separate `model' 
which may or may not fit with the global solutions of accretion and winds.

\section{Advective Flow Paradigm}

\subsection{Hydrodynamic and magnetohydrodynamic flows}

Though most of the literature concentrates on disks and jets separately, it is 
advisable (and {\it economical}) to study them simultaneously. This is because 
in global studies, always both types of solutions appear together. Examples 
are Bondi (1952) and generalized Bondi solutions (Chakrabarti, 1990; 1996bc). 
Numerical simulations have also shown outflows in much of the 
region of the parameter space (Molteni, Lanzafame \& Chakrabarti, 1994 [MLC94];
Molteni, Ryu \& Chakrabarti, 1996 [MRC96]; Ryu, Chakrabarti \& Molteni, 1997 [RCM97]).
Figure 1 shows the broad classification of all possible steady, inviscid 
solutions of advective disks (adapted from Chakrabarti, 2000a). Solutions 
shown are obtained for inviscid adiabatic flows (Chakrabarti 1989, 1996b)
but they remain valid close to the black hole even when these conditions 
are relaxed. One important aspect which requires emphasizing is the presence 
of a centrifugal pressure supported boundary layer (CENBOL) in some of these
steady solutions and in most of the non-steady solutions! Rapid inflow
keeps the angular momentum to an almost constant value. Thus, centrifugal
force becomes a dominant force. It slows downs matter in this barrier,
heats them up so that hard X-rays could come out and finally drives hot matter
perpendicular to the disk to form outflows. The barrier even oscillates
with a large amplitude if conditions are right (Molteni, Sponholz \&
Chakrabarti, 1996; RCM97) thereby giving rise to
quasi-periodic oscillations of X-rays. Beside each solution in Fig. 1 
schematic picture is given of accretion and outflows. Inward and outward pointing arrows 
indicate accretion wind solutions respectively. 
In (A) and (a), low energy and low angular momentum 
flow behaves like a conical Bondi inflow and Parker type winds 
well known in solar physics. In (B) and (b), a steady shock does not form, 
the inflow becomes hotter due to slowing down at the centrifugal barrier. 
Numerical simulation shows that an oscillating shock forms (RCM97)
which is accompanied by a non-steady outflow. In (C) and (c), the steady 
inflow solution has a standing shock. This has been tested by numerical 
simulations (Chakrabarti \& Molteni, 1993 [CM93]; MLC94; MRC96). Positive energy outflow
is also found to be steady. In (D) and (d), the inflow passes through an
inner sonic point, but the outflow forms a steady shock (CM93).
In (E) and (e), no shock is possible in the 
steady solution though there are three sonic points and we
feel that {\it non-steady} shocks would form. In (F) and (f),
both the inflow and the outflow have no shocks, and energetic steady solutions
are present as both pass through the inner sonic point. In (G) and (H), solutions
are incomplete since they are not from infinity to the horizon.
Thus, no steady solution is possible. Simulations (Ryu and Chakrabarti, 1996)
indicated that flow is highly unsteady in these cases which are 
schematically shown in panels (g) and (h) respectively. Especially interesting
is the solution of O*, where disks could be `thick' not because of gas or 
radiation pressure, but because of turbulent pressure. 
Solutions from (g) are with negative energy. 

When viscosity is present, closed topologies open up to join with 
Keplerian disks and when heating (such as magnetic, viscous)
flow from a Keplerian disk can pass through shocks and sonic points
(Chakrabarti 1990, 1996, 2000a). For instance, solutions in (C) 
become similar to those of (I) and (i) while those in I$^*$ become similar to those 
of (J) and (j). These solutions are the backbones of the advective 
disk paradigm and are useful in obtaining complete solutions 
of the disk/jet systems. When magnetic field is added, 
the flow passes through the fast and the slow magnetosonic
point and the Alf\'ven point. Possible outflowing solutions
with or without shocks have been presented in Chakrabarti (1990).
It is likely that in some regions of the parameter space,
these magnetohydrodynamic shocks should also show oscillations.

\begin{figure}
\vbox{
\vskip 0.5cm
\hskip 0.0cm
\centerline{
\psfig{figure=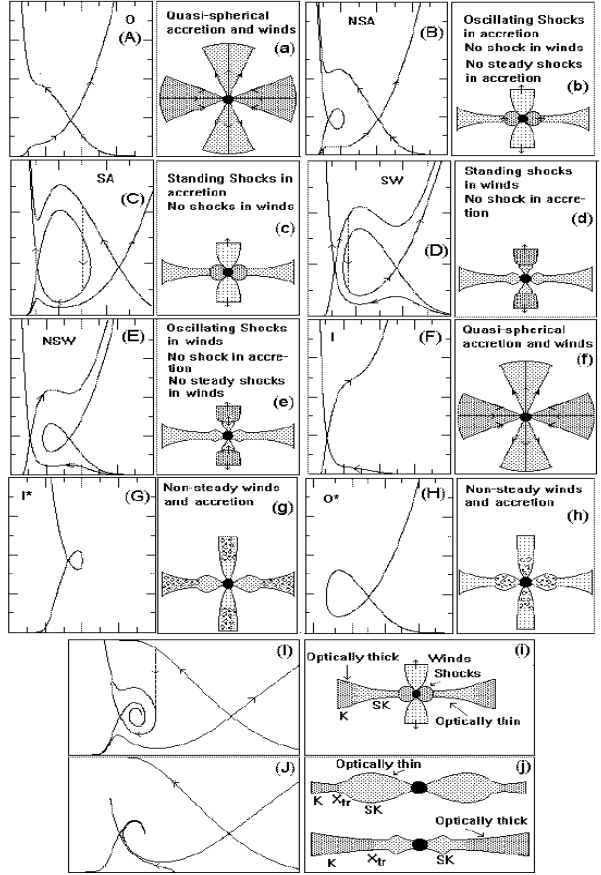,height=12truecm,width=8truecm}}}
\begin{verse}
\vspace{-1.0cm}
\caption[] {Nature of the solutions (Mach number along Y-direction and
logarithmic radial distance along X-direction) of inviscid (A-H) and 
viscous (I-J) advective disks. Schematic flow behaviour of these solutions 
in meridional plane are shown in (a-j). Shock locations (C, D) and incomplete solutions
(G, H) are indicated by puffed up (c, d) and turbulent regions (g, h) respectively. Solutions
with three sonic points having no steady shock are shown oscillating
with arrows at both ends (b, e).}
\end{verse}
\end{figure}

\subsection{Effects of Radiative Transfer}

Fully self-consistent two temperature flow in the advective disks 
was studied very recently. Chakrabarti \& Titarchuk (1995) computed 
temperatures of electrons and ions for various accretion rate parameters
in the CENBOL region and found that for high Keplerian rate CENBOL
cools down completely and shock disappears. Soft photons intercepted by CENBOL
are reprocessed by hot electrons to produce hard X-rays which are observed 
as a power-law tail in the hard  states. In soft states, 
the hard tail is formed due to bulk motion Comptonization. 

Hot CENBOL drives outflows from the disk along the vertical axis. When the outflow
rate is very high, the sonic sphere (i.e., region of the wind till its first 
sonic surface) becomes dense enough to cool down due to Comptonization
by the soft photons from the Keplerian disk. Estimation of the rate of wind generation
(Chakrabarti 1998ab, Chakrabarti 1999a) shows that in purely soft states
no outflow is possible and in purely hard states, outflow is very small. Thus above mentioned
cooling of the sonic sphere is not possible. When compression ratio of the 
shock is intermediate, outflow rate is significant.

After the sonic sphere is cooled sonic surface comes closer to the black hole
and matter below it returns back to the disk, while matter above it separates as
blobs. Thus blobby jets are expected in this intermediate states (this may also be called
flare/quiescence or On/Off transition states). The return flow acts as a 
feedback on the already accreting flow and the count rate undergoes very interesting
behaviour. Fig. 2 (Chakrabarti, 2000b) shows how the ratio ($R_{\dot m}$)
of the outflow rate ${\dot M}_{out}$ and the inflow rate (${\dot M}_{in}$ assumed to be proportional
to $1/R^2$; $R\rightarrow 1\rightarrow$ soft state; $R\rightarrow 7 \rightarrow$ hard state) 
depend on the compression ratio $R$ of the gas at the shock. This clearly shows that
there is a distinct relation between the spectral states and the outflow rate.

\begin{figure}
\vbox{
\vskip -2.5cm
\hskip -0.0cm
\centerline{
\psfig{figure=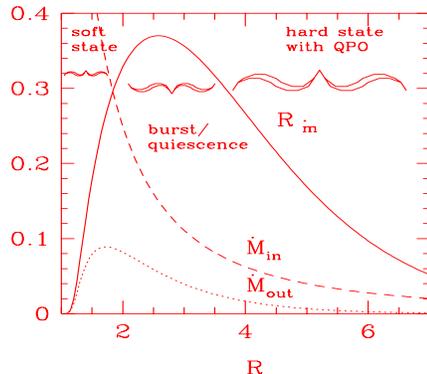,height=10truecm,width=10truecm}}}
\begin{verse}
\vspace{-3.0cm}
\caption[] {Variation of ratio of outflow and inflow rates
as a function of the compression ratio at the accretion shock. Also shown are
expected outflow rate (short-dashed curve) for a given inflow rate (long-dashed curve).}
\end{verse}
\end{figure}

An important characteristics the spectrum should show is that spectral slops 
of the power-law component should be harder in soft states and softer in the hard states.
This is because, in the hard states, outflows from CENBOL reduced electron density 
but the  soft-photon intensity from the pre-CENBOL flow remains the 
same. This softens the spectra in hard states (Chakrabarti, 1998c). Similarly return flow
hardens the spectrum. A corollary of these effects is that
the pivotal point of the power-law tail shifts to a larger energy in presence of winds/return 
flows. 

\section{Explanations of Observational Aspects of GRS1915+105}

GRS1915+105 exhibits very intriguing light curves. Belloni
et al. (2000, hereafter B00) divided them into twelve types. They draw HR1 vs HR2 diagrams
where HR1=B/A and HR2=C/A (A:2-5keV, B:5-13keV, C:13-60keV). Color-Color diagrams
showed very intricate structures (shapes of atoll, banana, etc.). If the pre-shock flow is 
indeed the source of the soft photons, photons originating in (0-3) keV 
range should be roughly proportional to the Keplerian accretion rate. 
Thus, photon number may show time variation (due to periodic changes in the `accretion
rate' due to return flow mentioned above). However, no QPO should be seen from these
photons. Chakrabarti \& Manickam (2000, hereafter CM00) demonstrated this (see also, Rao et al. 2000). 
The harder photons ($E>3$ keV) would usually come from the post-shock flow. Since the spectra intersect 
at around $17$keV, and for $E>17$keV, photon number is not large, we make our 
choice of A, B and C to be in ranges of $(0-3)$ keV, $(3-17)$ keV and $(17-60)$ keV
respectively (B and C would be related to sub-Keplerian rate).
According to our paradigm, roughly speaking, A, B and C should 
be proportional to each other (since B and C are produced by interception of soft 
photons measuring A. Of course, soft X-ray absorption makes matter more complex.)
and whenever hardness or softness ratios are plotted basically straight lines are 
expected, instead of Atoll, Banana and Z shapes.  The results are presented in
Nandi et al. (2000a) and the details of the physical interpretations are
presented in Manickam et al. (2000). 

One important conclusion of Belloni et al. (2000) was that the disk apparently
has three states: A (low rate and low HR1, HR2), B (high rate, high HR1) and 
C (low rate, low HR1, variable HR2 depending on length of the event). It seems that the State C exhibits
QPO. State A and state B do not exhibit QPO. More interestingly, except for
$C\rightarrow B$ transition, all other transitions of states are allowed. Nandi et al (2000b)
found evidence of QPO in some of the State A light curve.

As discussed in Chakrabarti (2000ab) there are practically four distinct ways that the solutions
in Fig. 1 could be combined for a realistic accretion-wind system.
If the accretion rate is very low and shock does not develop, QPO may not be seen. 
In GRS 1915+105 we need not be concerned with this type of flow. The remaining three are shown in Fig. 3
(Chakrabarti et al. 2000a). 
If the shock develops, QPO would be seen but wind may be negligible 
or absent depending on the compression ratio of the shock. The object would be in 
State C of B00 or hard class of Nandi et al. (2000). 
If the accretion rate is generally increased, shock is weakened (compression ratio goes down), 
and in the intermediate state, burst/quiescence or On/Off may be seen. 
There could be two types of `On' states (State A and State B of B00).
After the winds of State C fills in the sonic sphere and cools it
down by Comptonization, CENBOL and the region till the sonic sphere collapse. This is the State A
of B00. Now there are two possibilities (Chakrabarti, 1999b) either the flow 
separates completely as a blob and returns to State C or the flow mostly 
return back to the accretion disk and enhances the accretion.  
This would be the State B of B00. This may in turn increases the outflow 
(see, Fig. 2). But shock itself is getting weakened because of post-shock cooling, 
hence the outflow is very mild, but may be in a threshold so that a bit more outflow can cause 
the sonic sphere to collapse again. Thus, occasional trips to State A from State B is possible.
Once enhanced matter is drained out and the shock bounces back to roughly
the original location (compatible with its specific energy and angular momentum) 
State C forms again. Since State C produce fewer soft photons, State B is
not directly possible from State C without first producing return flow and 
enhanced accretion. This may explain why a transition from C to B is difficult.

\begin{figure}
\vbox{
\vskip 0.0cm
\hskip -0.0cm
\centerline{
\psfig{figure=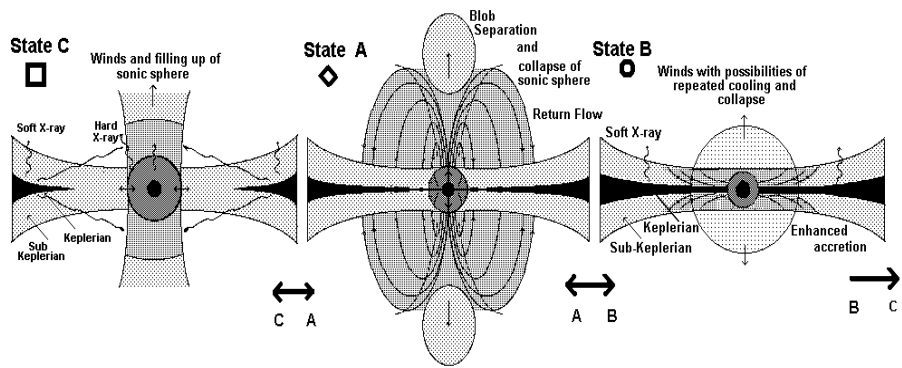,height=5truecm,width=12truecm}}}
\begin{verse}
\vspace{-1.0cm}
\caption[] {Suggested accretion/wind configurations in States A, B and C of B00.}
\end{verse}
\end{figure}

Among other confirmations of our paradigm, we note that
Nandi et al. (2000b) demonstrated that the spectral slopes of GRS 1915+105 are hardened
by enhanced accretion and softened by return flows.
CM00 found that the duration of the Off states strongly depend on QPO frequency.
Dhawan et al (2000) observationally demonstrates that jets originate from what we term CENBOL region.

\section{Concluding remarks}

Advective disk paradigm being made up of the most complete accretion disk/wind solutions, it
is not surprising that most of the observations could be explained by this paradigm. 
However, our explanations of the light curves of GRS 1915+105
have been over simplified since we ignored the effects of magnetic
fields altogether. Apart from a new spectral component due to synchrotron radiation
and perhaps increasing the outflow velocity to a much larger value, 
no other effect is expected as far as our State description is concerned. 

This work is partly supported by a grant from ISRO for the project `Quasi-Periodic Oscillations
of Black Holes'.

\end{article}
\end{document}